\documentclass[11pt]{article}
\usepackage{curves}
\def\citenum#1{{\def\@cite##1##2{##1}\cite{#1}}}
\def\citea#1{\@cite{#1}{}}
 
\def\eqs#1#2{{eqs. \ref{#1}--\ref{#2}}}

\newcommand{\be}{\begin{equation}}
\newcommand{\ee}{\end{equation}\noindent}
\newcommand{\bear}{\begin{eqnarray}}
\newcommand{\ear}{\end{eqnarray}\noindent}
\newcommand{\no}{\noindent}
\date{}
\renewcommand{\theequation}{\arabic{section}.\arabic{equation}}

\def\Eins{\mathord{1\hskip -1.5pt
\vrule width .5pt height 7.75pt depth -.2pt \hskip -1.2pt
\vrule width 2.5pt height .3pt depth -.05pt \hskip 1.5pt}}

\newcommand{\slD}{\raise.15ex\hbox{$/$}\kern-.57em\hbox{$D$}}
\newcommand{\slpartial}{\raise.15ex\hbox{$/$}\kern-.57em\hbox{$\partial$}}
\newcommand{\slG}{{{\dot G}\!\!\!\! \raise.15ex\hbox {/}}}

\def\GBd12{{\dot G}_{B12}}

\def\non{\nonumber}
\def\beqn*{\begin{eqnarray*}}
\def\eqn*{\end{eqnarray*}}

\def\square{\kern1pt\vbox{\hrule height 1.2pt\hbox{\vrule width 1.2pt
   \hskip 3pt\vbox{\vskip 6pt}\hskip 3pt\vrule width 0.6pt}
   \hrule height 0.6pt}\kern1pt}
\def\slash#1{#1\!\!\!\raise.15ex\hbox {/}}

\def\dps{\displaystyle}

\def\half{{1\over 2}}

\def\fourth{{1\over4}}

\def\4piTD{{(4\pi T)}^{-{D\over 2}}}
\def\4piT4{{(4\pi T)}^{-2}}

\def\Tintm4{{\dps\int_{0}^{\infty}}{dT\over T}\,e^{-m^2T}
    {(4\pi T)}^{-2}}
\def\Tintm{{\dps\int_{0}^{\infty}}{dT\over T}\,e^{-m^2T}}

\def\tr{{\rm tr}\,}

\def\bbbz{{\mathchoice {\hbox{$\sf\textstyle Z\kern-0.4em Z$}}
{\hbox{$\sf\textstyle Z\kern-0.4em Z$}}
{\hbox{$\sf\scriptstyle Z\kern-0.3em Z$}}
{\hbox{$\sf\scriptscriptstyle Z\kern-0.2em Z$}}}}
\newfont{\bbbold}{msbm10 scaled \magstep1}

\newfont{\goth}{eufm10 scaled \magstep1}

\def\S{\Sigma}

\def\be{\begin{equation}}\def\ee{\end{equation}}
\def\bea{\begin{eqnarray}}\def\eea{\end{eqnarray}}
\def\ba{\begin{array}}\def\ea{\end{array}}

\def\bd{\begin{document}}
\def\ed{\end{document}}
\def\bea{\begin{eqnarray}}
\def\eea{\end{eqnarray}}
\def\ba{\begin{array}}
\def\ea{\end{array}}
\def\ft#1#2{{\textstyle{{\scriptstyle #1}\over {\scriptstyle #2}}}}
\def\fft#1#2{{#1 \over #2}}
\def\eqs#1#2{(\ref{#1}-\ref{#2})}
\def\det{{\rm det\,}}
\def\tr{{\rm tr}}
\newcommand{\ho}[1]{$\, ^{#1}$}
\newcommand{\hoch}[1]{$\, ^{#1}$}
\newcommand{\tamphys}{\it\small Center for Theoretical Physics,
Texas A\&M University, College Station, TX 77843, USA} 
\newcommand{\newton}{\it\small Isaac Newton Institute for Mathematical
Sciences, Cambridge, UK} 
\newcommand{\kings}{\it\small Department of Mathematics, King's College,
London, UK} 
\newcommand{\lapp}{\it\small LAPP, Annecy, France}

\newcommand{\auth}{\large B. Eden,
C. Schubert, E. Sokatchev} 
\begin{document}\input feynman

\thispagestyle{empty} 

\hfill{LAPTH-786/2000} 

\hfill{\today}

\vspace{40pt} 

\begin{center}
{\Large{\bf
Three-Loop Four-Point Correlator in $N=4$ SYM
}} 

\vspace{50pt} 

\auth

\vspace{15pt} 

{\it Laboratoire d'Annecy-le-Vieux de Physique
Th{\'e}orique\footnote{UMR 5108 associ{\'e}e {\`a} 
l'Universit{\'e} de Savoie} LAPTH, \\
Chemin de Bellevue, B.P. 110, \\
F-74941 Annecy-le-Vieux, France} 

\vspace{100pt} 
{\bf Abstract} 

\end{center}
We explicitly compute the complete three-loop (O($g^4$)) 
contribution to the four-point function of chiral primary 
current-like operators $\langle\tilde q^2 q^2\tilde q^2 
q^2\rangle$ in any finite $N=2$ SYM theory.
The computation uses $N=2$ harmonic supergraphs in 
coordinate space. Dramatic simplifications are achieved by a 
double insertion of the $N=2$ SYM linearized action, and 
application of superconformal covariance arguments to the 
resulting nilpotent six-point amplitude. The result involves 
polylogarithms up to fourth order of the conformal cross ratios. 
It becomes particularly simple in the $N=4$ special case.

\pagebreak \setcounter{page}{1}
\renewcommand{\theequation}{\arabic{equation}}
\setcounter{equation}{0}
\vskip10pt

Correlators of gauge invariant composite operators are natural 
objects to study in finite super-Yang-Mills theories, since they 
are strongly constrained by superconformal invariance. These 
constraints have recently been actively investigated using both 
abstract methods 
\cite{hw1}-\cite{ehpsw} and the operator product expansion 
\cite{osborn,intski,bkrs}. Additional motivation is provided 
by the AdS/CFT correspondence conjecture \cite{m,gkp,witten}, 
which relates correlators of chiral primary operators in $N=4$ 
super-Yang-Mills theory to correlators in AdS supergravity 
\cite{muevis}-\cite{psz}. This relation involves the SYM 
correlators at strong coupling, so that its verification presently 
relies mostly (although not exclusively \cite{colorful}) on 
non-renormalization theorems 
\cite{dfs},\cite{gubkle}-\cite{erdvic}. 

On the CFT side, the correlator of four $N=4$ stress tensors is 
the simplest one which can be built from chiral primary operators, 
and is {\sl not} subject to any known non-renormalization theorem. 
The quantum corrections to this correlator have so far been 
investigated to lowest-order in the perturbative (two-loop) 
\cite{ehssw1}-\cite{grps} and non-perturbative (one-instanton) 
\cite{colorful} sectors. From the point of view of the operator
product expansion approach it is also of interest to
know the singularity structure of this correlator
at coincidence points \cite{bkrs,dmmr}. 

In this paper we a make a step further by 
computing it at the next, three-loop level. Our computation is not 
restricted to $N=4$ SYM theory but concerns the correlator of four 
bilinear (current-like) hypermultiplet matter composite operators 
in $N=2$ SYM. There are three reasons for staying at the $N=2$ 
level: i) there is no known off-shell formulation of $N=4$ SYM, so 
the best way to carry out a quantum calculation is to reformulate 
the theory in $N=2$ harmonic superspace \cite{gikos1} and then use 
the efficient supergraph technique available there \cite{gikos2}; 
ii) as shown in \cite{ehssw1}, knowing the result in the $N=2$ 
matter sector and using the $SU(4)$ R symmetry of the $N=4$ theory 
one can easily reconstruct the complete amplitude for four $N=4$ 
stress tensors; iii) it is not impossible that many of the 
expected exceptional features of four-dimensional CFT are shared 
by all finite $N=2$ theories \cite{fin1,fin2}.  

The correlator we consider is made out of hypermultiplets 
$q^+,\tilde q^+$ and has the form 

\bear G &=& \langle{\rm tr}(\tilde q^2) 
{\rm tr}(q^2){\rm tr}(\tilde q^2) 
{\rm tr}(q^2)\rangle \equiv 
\langle{\cal O}_1\cdots{\cal O}_4 \rangle \label{defG} \ear\no 

In the $SU(2)$-covariant harmonic superspace formalism 
\cite{gikos1} this hypermultiplet is described off shell by a 
Grassmann analytic superfield 
$q^+(x_A,\theta^+,\bar\theta^+,u^{\pm})$. The harmonic variables are 
defined as $SU(2)$ matrices, 

\begin{equation}
 u\in SU(2)\quad \Rightarrow\quad   u^-_i =(u^{+i})^*\;, 
\quad u^{+i}u^-_i = 1 
\label{1'}  
\end{equation}\no
The Grassmann variables  

\begin{equation}
  \theta^{+\alpha} = u^+_i\theta^{i\alpha},\quad 
\bar\theta^{+\dot\alpha} = u^+_i\bar\theta^{i\dot\alpha}  
\label{1''}
\end{equation}
are $SU(2)$-invariant $U(1)$ projections of the full superspace 
ones $\theta^i,\bar\theta^i$. The coordinates 
$x^{\alpha\dot\alpha}_A = x^{\alpha\dot\alpha} -4i \theta^{(i\, 
\alpha} 
  \bar\theta^{j)\, \dot\alpha}u^+_iu^-_j$ together with 
$\theta^+,\bar\theta^+$ and $u^\pm$ span the G-analytic superspace 
closed under the full $N=2$ superconformal group (see 
\cite{gikos1,fradkin} for details). In order for the theory to be 
finite, the matter hypermultiplets $q^+$ must be in a 
representation $r$ of the gauge group such that 
$C(r)= C_2(G)$ \cite{fin1,fin2}
\footnote{We denote the generators in this representation
by $t^a$, and ${\rm tr}(t^at^b)= C(r)\delta^{ab}$,
$t^at^a = C_2(r)\cdot\Eins$.}. 

A direct calculation of this correlator at three loops using 
standard component or $N=1$ superfield techniques would be, at the 
present state-of-the-art, a prohibitively difficult task. In the 
following we will perform it in a rather roundabout way, using the 
new set of coordinate-space Feynman rules for $N=2$ harmonic 
superspace given in \cite{hssw}, together with a fortuitous 
combination of Intriligator's insertion trick and knowledge about 
the construction of superconformal invariants which was built up 
in \cite{hsw,ehw,ehssw1,ehssw2,hssw,ehssw3,ehpsw}. 

Intriligator's trick, introduced in the present context in 
\cite{intriligator}, allows us to write the three-loop (O($g^4$)) 
contribution to this correlator in terms of a double insertion of 
the linearized SYM action, 

\bear G^{\rm 3-loop} &\sim& \int d^4x_5 d^4\theta_5 \int d^4x_6 
d^4\theta_6 \Bigl\langle {\cal O}_1\cdots{\cal O}_4 {\tr} W_{5}^2 
{\tr} W_{6}^2 \Bigr\rangle \non\\ \label{intriligation} \ear\no 
where the integrals are over $N=2$ chiral superspace. This 
representation can be derived either by a simple path integral 
manipulation \cite{intriligator,hssw}, or diagrammatically by 
showing that this expression differs from the original set of 
Feynman diagrams only by a simultaneous change of gauge for all 
SYM propagators \cite{ess}. One advantage of this representation 
is that it leads, due to the fact that $W$ is chiral and $q^+$ is 
G-analytic, to severe restrictions on the possibilities for 
building superconformal invariants. 

We have already successfully employed this trick in the analogous 
two-loop calculation.  There one makes a single insertion and 
deals with the five-point correlator $\langle {\cal 
O}_1\cdots{\cal O}_4 {\tr} W_5^2 \rangle$. It is easy to see that 
superconformal invariance alone completely determines its 
Grassmann dependence \cite{ehssw3,ehpsw}. Indeed, the correlator 
must be a superconformal covariant having the R weight of four 
left-handed $\theta$'s (in order to mach that of the chiral 
measure $d^4x_5d^4\theta_5$). This means it has to be made out of 
combinations of $\theta^+_{1,\ldots,4}$ and $\theta^i_5$ invariant 
under the shift part of the superconformal transformations. Since 
the latter involve 4 left-handed odd parameters, there exist only 
two such combinations $\xi_{123}$ and $\xi_{124}$ given below 
\cite{osborn,ehw,ehssw3}: 

\bear \xi_{12m}^{\dot\alpha} &\equiv& (12)\rho_m^{\dot\alpha} + 
(2m)\rho_1^{\dot\alpha} +(m1)\rho_2^{\dot\alpha}, \quad m=3,4, 
\non\\ \rho_a^{\dot\alpha} &\equiv& 
\theta_{5a\alpha}{x^{\alpha\dot\alpha}_{5a}\over x_{5a}^2}, \quad 
a=1,\ldots ,4, \non\\ \theta_{5a}^{\alpha} &\equiv& 
u_{ai}^{+}\theta_5^{i\alpha}-\theta_a^{+\alpha} \non\\ 
\label{defxi,rho,thetadiff} \ear\no ($(12) = u_1^{+i}u_{2i}^+,\, 
x_{5a} = x_{L5}-x_{Aa}$).   Then it is clear that the correlator 
must have the form 

\bear \Bigl\langle {\cal O}_1\cdots{\cal O}_4 {\tr} W_5^2 
\Bigr\rangle &=& \xi^4 F(x,u) + {\rm O}(\theta^5\bar\theta) 
\label{2loopform} \ear\no where $\xi^4 \equiv (12)^{-2} 
{\xi_{123}^2\xi_{124}^2}$, and the factor $F(x,u)$ is a conformal 
covariant depending only on the space-time and harmonic variables. 
The latter {\it is not predicted by invariance alone} and had to 
be determined by an explicit graph calculation. It was rather 
surprising to find out that the result was a {\it rational 
function} of the space-time variables. Then, after the single 
integration over the insertion point $\int d^4x_5$ we found that 
the entire amplitude was expressed in terms of the one-loop scalar 
box integral. 

Extending the same argument based on counting the R weight and the 
number of independent shift-invariant combinations of left-handed 
$\theta$'s to the three-loop case, one finds that superconformal 
invariance constrains the integrand to be of the form 

\bear \Bigl\langle {\cal O}_1\cdots{\cal O}_4 {\tr} W_5^2{\tr} 
W_6^2 \Bigr\rangle &=& \xi^4\psi^4 G(x,u) + {\rm 
O}(\theta^9\bar\theta) \label{3loopform} \ear\no where $\xi^4$ is 
the same as above, and $\psi^4$ the corresponding covariant 
referring to point `6', built from $\sigma_a^{\dot\alpha}\equiv 
\theta_{6a\alpha}{x^{\alpha\dot\alpha}_{6a}\over x_{6a}^2}$. Once 
again, the purpose of the graph calculation is to determine the 
factor $G(x,u)$. 

The knowledge of the dependence on the odd coordinates in 
(\ref{3loopform}) is extremely useful, since it allows one to 
concentrate on one typical term in the expansion of the nilpotent 
covariant while doing the explicit graph calculation. Thus, in the 
two-loop calculation \cite{hssw} a very substantial simplification 
was reached by setting the analytic Grassmann variables to zero, 
$\theta_1^+=\ldots =\theta_4^+ =0$, and keeping only the chiral 
$\theta_5$. At the three-loop level, it turns out that even more 
dramatic simplifications can be achieved by keeping only the 
external Grassmann variables, and setting to zero both $\theta_5$ 
and $\theta_6$. Further, we are only interested in the leading 
term in (\ref{3loopform}), so we can set all right-handed 
$\bar\theta^+_{1,\ldots,4}=0$. Then 
the whole correlator becomes  proportional to 
$\theta_1^{+2}\theta_2^{+2} \theta_3^{+2}\theta_4^{+2}$. Now,  
this nilpotent factor absorbs the complete U(1) charge, so its 
coefficient is chargeless and hence {harmonic independent} 
\footnote{This argument is based on harmonic analyticity. It 
should be stressed that the latter is not an assumption but can 
easily be proven by examining the complete set of 3-loop graphs 
\cite{ess}.}. This allows one to identify all harmonic variables, 
$u_1 = \ldots = u_4$.  The combination of all this turns out to 
have the effect of eliminating all Feynman diagrams except those 
with exactly one interaction vertex along every matter line. Up to 
permutations, this leaves only the three diagrams depicted in fig. 
1. (In particular, all diagrams involving gauge self-interactions 
drop out.) 
\vspace{20pt}
\begin{center}
\begin{picture}(42000,15000)(0,-5000)

\drawline\gluon[\E\REG](0,4100)[10] \global\Xone = \gluonbackx 
\drawline\gluon[\S\REG](6660,\Xone)[5] 
\drawline\gluon[\N\FLIPPED](6660,0)[2] 
\drawline\fermion[\E\REG](0,0)[\Xone] 
\drawarrow[\W\ATTIP](\pmidx,\pmidy) 
\drawline\fermion[\N\REG](0,0)[\Xone] 
\drawarrow[\S\ATTIP](\pmidx,\pmidy) 
\drawline\fermion[\W\REG](\Xone,\Xone)[\Xone] 
\drawarrow[\E\ATTIP](\pmidx,\pmidy) 
\drawline\fermion[\S\REG](\Xone,\Xone)[\Xone] 
\drawarrow[\N\ATTIP](\pmidx,\pmidy) \put(6660,7540){\circle*{500}} 
\put(6660,7270){\circle*{500}} \put(7160,7170){5} 
\put(3350,4100){\circle*{500}} \put(3080,4100){\circle*{500}} 
\put(3060,4680){6} \put(5000,-3000){A} 

\drawline\fermion[\E\REG](13000,0)[\Xone] \global\advance\pmidx by 
-400 \drawarrow[\W\ATTIP](\pmidx,\pmidy) 
\drawline\fermion[\N\REG](13000,0)[\Xone] \global\advance\pmidy by 
-400 \drawarrow[\S\ATTIP](\pmidx,\pmidy) \global\Xtwo = 13000 
\global\advance\Xtwo by \Xone 
\drawline\fermion[\W\REG](\Xtwo,\Xone)[\Xone] 
\global\advance\pmidx by 400 \drawarrow[\E\ATTIP](\pmidx,\pmidy) 
\drawline\fermion[\S\REG](\Xtwo,\Xone)[\Xone] 
\global\advance\pmidy by 400 \drawarrow[\N\ATTIP](\pmidx,\pmidy) 
\drawline\gluon[\SE\FLIPPED](17160,\Xone)[6] \global\Ythree = 
\Xone \global\advance\Ythree by -3300 \global\Xthree = \Xone 
\global\advance\Xthree by 13000 \global\advance\Xthree by -3300 
\global\advance\Xthree by -250 
\put(\Xthree,\Ythree){\circle*{500}} \global\advance\Xthree by 250 
\global\advance\Ythree by -250 
\put(\Xthree,\Ythree){\circle*{500}} \global\advance\Xthree by 
-600 \put(\Xthree,5900){5} \drawline\gluon[\SE\REG](13000,6500)[6] 
\global\Xthree = 16300 \global\Ythree = 3300 
\global\advance\Xthree by 250 \put(\Xthree,\Ythree){\circle*{500}} 
\global\advance\Xthree by -250 \global\advance\Ythree by 250 
\put(\Xthree,\Ythree){\circle*{500}} \global\advance\Xthree by 120 
\put(\Xthree,4100){6} \put(18000,-3000){B} 

\startphantom \drawline\fermion[\E\REG](26000,0)[\Xone] 
\stopphantom \global\Xfour = \pmidx \global\Xfive = \pbackx 
\global\Yfour = 1045 \global\Yfive = \Yfour \global\advance\Yfive 
by \Yfour \curve(26000,\Yfour,\Xfour,0,\Xfive,\Yfour) 
\curve(26000,\Yfour,\Xfour,\Yfive,\Xfive,\Yfour) 
\global\advance\Xfour by -200 \drawarrow[\W\ATTIP](\Xfour,0) 
\drawarrow[\W\ATTIP](\Xfour,\Yfive) \global\advance\Xfour by 200 
\global\Ysix = \Xone \global\advance\Ysix by -\Yfour 
\global\Yseven = \Ysix \global\advance\Yseven by -\Yfour 
\curve(26000,\Ysix,\Xfour,\Xone,\Xfive,\Ysix) 
\curve(26000,\Ysix,\Xfour,\Yseven,\Xfive,\Ysix) 
\global\advance\Xfour by 200 \drawarrow[\E\ATTIP](\Xfour,\Xone) 
\drawarrow[\E\ATTIP](\Xfour,\Yseven) \global\advance\Xfour by -200 
\global\Yone = 220 \global\Yeight = \Xone \global\advance\Yeight 
by -\Yone \drawline\gluon[\S\FLIPPED](29000,\Yeight)[8] 
\global\advance\Xfour by -26000 \global\advance\Xfour by -270 
\global\advance\Xfour by -135 \global\advance\Xfour by -200 
\put(27900,\Xfour){5} \global\advance\Xfour by 200 
\put(29000,\Xfour){\circle*{500}} \global\advance\Xfour by 270 
\put(29000,\Xfour){\circle*{500}} \global\advance\Xfive by -3000 
\drawline\gluon[\N\FLIPPED](\Xfive,\Yone)[8] \global\advance\Xfour 
by 540 \put(\Xfive,\Xfour){\circle*{500}} \global\advance\Xfour by 
-270 \put(\Xfive,\Xfour){\circle*{500}} \global\advance\Xfive by 
600 \global\advance\Xfour by -200 \put(\Xfive,\Xfour){6} 
\put(31000,-3000){C} 

\put(16500,-4500){Figure 1} 
\end{picture}
\end{center}
\no
\vspace{10pt}
These diagrams involve only the ``building block'' 
shown in fig. 2. 

\begin{center}
  \begin{picture}(0,3000)

  \drawline\gluon[\S\CENTRAL](0,0)[4]
  \put(\gluonbackx,\gluonbacky){\circle*{500}}
  \drawline\fermion[\W\REG](\gluonfrontx,\gluonfronty)[5000]
  \drawarrow[\E\ATTIP](\pmidx,\pmidy)
  \global\advance\pbackx by -1200
  \put(\pbackx,\pbacky) {1a}
  \drawline\fermion[\E\REG](\gluonfrontx,\gluonfronty)[5000]
  \drawarrow[\E\ATBASE](\pmidx,\pmidy)
  \global\advance\pbackx by 500
  \put(\pbackx,\pbacky) {2c}
  \global\advance\gluonbackx by 1000
  \global\advance\gluonbacky by -500
  \put(\gluonbackx,\gluonbacky) {5,6\,b}
  \global\advance\gluonfronty by 1000
  \end{picture}
  \end{center}
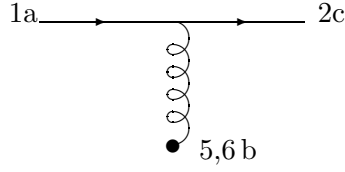
\vspace{20mm} \centerline{Figure 2} \vspace{5mm}

This building block, which we denote by $I_{5,6}$, is already 
known from the two-loop calculation \cite{hssw}. With the stated 
specializations, $\theta_{5,6}=\bar\theta_i^+=u_i-u_j=0$, the 
expression obtained there can be rewritten in terms of the 
variables $\rho_i (\sigma_i)$ as follows, 
%
%
\bear
I_{5}
&=&
{ig(t^b)^{ac}\over (2\pi)^4 x_{12}^2}
\,(\rho_1-\rho_2)^2
\label{I5simp}
\ear\no
Note that here the integration over the interaction
point has already been performed.
We can thus immediately write down
the contributions of all graphs to the six-point correlator.
The result reads, after some fierzing 
(up to an overall factor),

\bear
A &=& {C_A\over x_{12}^2x_{23}^2x_{34}^2x_{41}^2}
\Bigl\lbrack
\tau_{13}\tau_{24}-\tau_{12}\tau_{34}-\tau_{14}\tau_{23}
-\tau_{24}(\rho_1^2\sigma_3^2+\rho_3^2\sigma_1^2)
-\tau_{13}(\rho_2^2\sigma_4^2+\rho_4^2\sigma_2^2)
\non\\&&\quad\quad
+\tau_{12}(\rho_3^2\sigma_4^2+\rho_4^2\sigma_3^2)
+\tau_{34}(\rho_1^2\sigma_2^2+\rho_2^2\sigma_1^2)
+\tau_{14}(\rho_2^2\sigma_3^2+\rho_3^2\sigma_2^2)
+\tau_{23}(\rho_1^2\sigma_4^2+\rho_4^2\sigma_1^2)
\Bigr\rbrack
\non\\
B &=& {C_B\over x_{12}^2x_{23}^2x_{34}^2x_{41}^2}
\Bigl\lbrack
\tau_{13}(\rho_2^2\sigma_4^2+\rho_4^2\sigma_2^2)
+\tau_{24}(\rho_1^2\sigma_3^2+\rho_3^2\sigma_1^2)
\Bigr\rbrack\non\\
C &=&  C_C\Bigl\lbrack 
{\tau_{12}\tau_{34}\over x_{12}^4x_{34}^4}
+
{\tau_{14}\tau_{23}\over x_{14}^4x_{23}^4}
\Bigr\rbrack
\non\\
\label{ABC}
\ear\no
where
$\tau_{ij}\equiv 4(\rho_i\rho_j)(\sigma_i\sigma_j)
+\rho_i^2\sigma_j^2 +\rho_j^2\sigma_i^2$. The color factors 
$C_{A,B,C}$ are 

\bear C_A &=& d(G)C(r)\Bigl[C_2(r)-\half C_2(G)\Bigr]\non\\ C_B 
&=& d(G)C(r)C_2(r)\non\\ C_C &=& d(G)C^2(r)\non\\ 
\label{colourfactors} \ear\no
($d(G)$ denotes the dimension of the gauge group).
In terms of the variables 
$\tau_{ij}$ the six-point nilpotent covariant reads 

\bear \xi^4\psi^4 &=& (12)^2(34)^2\tau_{14}\tau_{23} + 
(14)^2(23)^2\tau_{12}\tau_{34} \non\\&& +(12)(23)(34)(41)\Bigl[ 
\tau_{13}\tau_{24}-\tau_{12}\tau_{34}-\tau_{14}\tau_{23}\Bigr] 
\label{xipsi} \ear\no For $\theta_5=\theta_6=0$ all these 
expressions are easy to evaluate, since then 

\bear \rho_i^2 &=& {\theta_i^{+2}\over x_{i5}^2}, \quad \sigma_i^2 
= {\theta_i^{+2}\over x_{i6}^2}, \quad \tau_{ij} = 
\theta_i^{+2}\theta_j^{+2}{x_{ij}^2x_{56}^2 \over 
x_{i5}^2x_{i6}^2x_{j5}^2x_{j6}^2} \label{rhotauexplicit} \ear\no 
Thus the six-point covariant becomes 

\bear \xi^4\psi^4\mid_{\theta_{5,6}=0} &=& 
\theta_1^{+2}\theta_2^{+2}\theta_3^{+2}\theta_4^{+2} {x_{56}^4 
R'\over \prod_{i=1}^4x_{i5}^2x_{i6}^2} \label{xipsianalytic} 
\ear\no where 

\bear
R' &=& (12)^2(34)^2 x_{14}^2x_{23}^2 +(14)^2(23)^2x_{12}^2x_{34}^2
\non\\&&
+(12)(23)(34)(41)\Bigl[
x_{13}^2x_{24}^2-x_{12}^2x_{34}^2-x_{14}^2x_{23}^2\Bigr]
\label{Rprime}
\ear\no
The coefficient function $G(x,u)$ can now be determined
by comparing the $\theta_1^{+2}\theta_2^{+2}\theta_3^{+2}\theta_4^{+2}$
 coefficients of the six-point correlator and the covariant.
Once this is done we know this correlator in covariant form, so 
that we can now return to the ``opposite'' frame where 
$\theta_1^+=\ldots =\theta_4^+ =0$. Here the covariant also looks 
very simple, 

\bear \xi^4\psi^4\mid_{\theta^+_{1,2,3,4}=0} &=& 
\theta_5^4\theta_6^4 {{R'}^2\over \prod_{i=1}^4x_{i5}^2x_{i6}^2} 
\label{xipsichiral} \ear\no 

It is remarkable and rather unexpected that this result only 
involves a {\it rational} function of the space-time variables, 
just like in the two-loop calculation. The Grassmann integrations 
being trivial ($\theta^4_{5,6}$ are just chiral delta functions), 
one is left with the space-time integrals $\int dx_5\int dx_6$. 
Only two different integrals appear, namely the standard one-loop 
box integral 

\bear h^{(1)}(x_1,x_2,x_3,x_4) \equiv  \int {dx_5\over 
x^2_{15}x^2_{25}x^2_{35}x^2_{45}} = -{i\pi^2\over 
x^2_{13}x^2_{24}} \Phi^{(1)} \Bigl({x_{12}^2x_{34}^2\over 
x_{13}^2x_{24}^2}, {x_{14}^2x_{23}^2\over x_{13}^2x_{24}^2}\Bigr) 
\label{defh} \ear\no and one other conformally invariant integral, 
e.g., 

\bear h_{12}^{(2)}(x_1,x_2,x_3,x_4) &\equiv& x_{12}^2 \int 
dx_5\int dx_6\, {1\over x_{15}^2x_{25}^2x_{35}^2x_{56}^2 
x_{16}^2x_{26}^2x_{46}^2} \non\\ && = {(i\pi^2)^2\over 
x_{12}^2x_{34}^2}\Phi^{(2)} \Bigl({x_{13}^2x_{24}^2\over 
x_{12}^2x_{34}^2}, {x_{14}^2x_{23}^2\over x_{12}^2x_{34}^2}\Bigr) 
\label{defi} \ear\no 
 The first one is well-known \cite{hv}, while 
the second one can be rewritten in terms of the two-loop (momentum 
space) double box integral, calculated in \cite{ussdavplb298}. The 
functions $\Phi^{(1,2)}$ are the first two elements of the 
infinite series of conformal ``multi-ladder'' functions introduced 
by Davydychev and Ussyukina \cite{ussdavplb305,broadhurst}. They 
can be written in terms of polylogarithms ${\rm Li}_n$ as follows 
\cite{ussdavplb298}, 

\bear
\Phi^{(1)}(x,y)
&=&
{1\over \lambda}
\Biggl\lbrace
2\Bigl({\rm Li}_2(-\rho x)
+
{\rm Li}_2(-\rho y)\Bigr)
+\ln
{y\over x}
\ln
{{1+\rho y}\over {1+\rho x}}
+
\ln (\rho x)\ln (\rho y)
+
{\pi^2\over 3}
\Biggr\rbrace
\non\\
\Phi^{(2)}(x,y)
&=&
{1\over \lambda}
\Biggl\lbrace
6\Bigl(
{\rm Li}_4 (-\rho x) + {\rm Li}_4 (-\rho y)
\Bigr)
+ 3\ln {y\over x}
\Bigl( {\rm Li}_3 (-\rho x) - {\rm Li}_3 (-\rho y)
\Bigr)
\non\\
&&
+ \half \ln^2 {y\over x}
\Bigl( {\rm Li}_2 (-\rho x) + {\rm Li}_2 (-\rho y) \Bigr)
+\fourth \ln^2 (\rho x)\ln^2(\rho y)
\non\\
&&
+\half \pi^2 \ln (\rho x)\ln (\rho y)
+ {1\over 12} \pi^2\ln^2 {y\over x} + {7\over 60} \pi^4
\Biggr\rbrace
\non\\
\label{Phi12explicit}
\ear\no
where
$
\lambda(x,y)= \sqrt{(1-x-y)^2-4xy},
\,
\rho(x,y)= 2(1-x-y+\lambda)^{-1}.
$
The final result after integration over points $5,6$ is reached by 
replacing, in eq. (\ref{ABC}),  

\bear
\tau_{12}\tau_{34} &\to& R'x_{12}^2x_{34}^2
\Bigl(h^{(1)}(x_1,x_2,x_3,x_4)\Bigr)^2, \non\\
\tau_{12}(\rho_3^2\sigma_4^2+\rho_4^2\sigma_3^2)
&\to&
2R'h_{12}^{(2)}(x_1,x_2,x_3,x_4)
\non\\
\label{replacerule}
\ear\no
etc. This result holds for any finite $N=2$ SYM theory. It considerably
simplifies if one specializes to the $N=4$ case, 
where $(t^b)^{ac}=if^{abc}$, and to the gauge group
$SU(N_c)$. Here all three contributions can
be added up, leading to

\bear\int_5\int_6 \Bigl\langle {\cal O}_1\cdots{\cal O}_4 {\tr} 
W_5^2 {\tr} W_6^2 \Bigr\rangle \mid_{\theta^+_{1,2,3,4}=0} &\sim& 
{(N_c^2-1)N_c^2 R'\over x_{12}^2x_{23}^2x_{34}^2x_{41}^2} \Bigl[ 
(x_{12}^2x_{34}^2+x_{13}^2x_{24}^2+x_{14}^2x_{23}^2) (h^{(1)})^2 
\non\\&&\hspace{80pt} +4(h_{12}^{(2)}+h_{13}^{(2)}+h_{14}^{(2)}) 
\Bigr] \label{finalN=4} \ear\no 

The details of this calculation will be given elsewhere 
\cite{ess}, as well as a discussion of the result. Here we only 
mention one important point. We have checked that our result 
exhibits a singularity of the type $\log^2 x^2_{12}$ in the 
coincidence limit $x_{12}\rightarrow 0$, exactly as predicted in 
\cite{bkrs,dmmr}.

It should also be mentioned that our three-loop result already 
dispenses with a speculation made in \cite{ehssw2}. There it was 
suggested that the unexpected absence of three- and 
quadrilogarithms in the two-loop result may be related to the fact 
that the complete tree-level result for the corresponding 
axion/dilation amplitudes in AdS supergravity can be represented 
in terms of logarithms and dilogarithms of the conformal cross 
ratios \cite{dfmmr}. With hindsight, the simplicity of the 
two-loop result is just a consequence of the fact that, apart from 
the standard box integral $h^{(1)}$, no other finite and 
conformally invariant scalar integral exists at this level. 

\vspace{15pt}\no {\bf Acknowledgements:} C.S. thanks A. Davydychev 
for detailed information on refs. 
\cite{ussdavplb298,ussdavplb305}. E.S. is grateful to S. Ferrara, 
R. Kallosh and R. Stora for enlightening discussions. C.S. and 
E.S. profited a lot from discussions with E. D'Hoker.

\vfill\eject

\end{document}